\documentclass[twocolumn,showpacs,amsmath,amssymb,superscriptaddress,prl]{revtex4}

\usepackage{graphicx,color,colordvi}
\usepackage{dcolumn}
\usepackage{bm}

\usepackage{times}

\begin{document}


\title{Well-localized edge states in two-dimensional topological
insulators: ultrathin Bi films }

\author{M. Wada}
\affiliation{Department of Physics, Tokyo Institute of Technology, 
Ookayama, Meguro-ku, Tokyo 152-8551, Japan} 
\author{S. Murakami}
\affiliation{Department of Physics, Tokyo Institute of Technology, 
Ookayama, Meguro-ku, Tokyo 152-8551, Japan} 
\affiliation{PRESTO, Japan Science and Technology Agency (JST),
Kawaguchi, Saitama 332-0012, Japan} 
\author{F. Freimuth}
\affiliation{Institut f\"{u}r Festk\"{o}rperforschung and Institute for Advanced Simulation, Forschungszentrum J\"{u}lich, D-52425 J\"{u}lich, Germany}
\author{G. Bihlmayer}
\affiliation{Institut f\"{u}r Festk\"{o}rperforschung and Institute for Advanced Simulation, Forschungszentrum J\"{u}lich, D-52425 J\"{u}lich, Germany}

\date{\today}

\begin{abstract}
We theoretically 
study the generic behavior of the penetration depth of the edge states 
 in two-dimensional quantum spin Hall systems.
We found that the momentum-space width of the edge-state dispersion 
 scales with the inverse of the 
penetration depth. As an example of well-localized edge states, 
we take the Bi(111) ultrathin film. Its edge states are found 
to extend almost over the whole Brillouin zone. Correspondingly, 
the bismuth (111) 1-bilayer system is proposed to have
 well-localized edge states 
in contrast to the HgTe quantum well.
\end{abstract}

\maketitle

{\it Introduction}-- The quantum spin Hall (QSH) phase \cite{Kane05a,Bernevig06a} is a new state of matter predicted theoretically,
and has received a lot of attention recently.
This phase is a nonmagnetic insulator in the bulk or film, and has gapless surface or edge states.
The edge states consist of counterpropagating states with opposite spins.
The notable feature of these edge states is that they are topologically protected;
they remain gapless even in the presence of nonmagnetic impurities and interaction \cite{Wu06,Xu06a}.
We still know few systems in which the QSH phase is realized.
The first theoretical proposal for the QSH system on the 
Bi ultrathin film by
one of the authors \cite{Murakami06a}. In addition, HgTe quantum well has been 
theoretically proposed \cite{Bernevig07a}, and experimentally 
shown to be in the 2D QSH state 
\cite{Konig07,Roth09}.

The edge states are localized near the edge, but 
their penetration depth 
$\ell$ into the bulk varies between the 
systems. The observation and control of 
the edge states crucially depends on the penetration depth, and it is an 
important issue how they are determined in various systems.
In the present paper we study the 
behavior of penetration depth $\ell$ in QSH systems. From a simple model 
we show that the minimum penetration depth  (which is typically reached 
in the middle of the bulk gap)  scales with 
the inverse of the extension of the edge states in  $k$-space.
Therefore, if the edge states exist only in a small region in $k$-space,
$\ell$ is long. 
By extending this conclusion to generic cases, we expect that the penetration 
depth is of the order of the lattice constant, if the edge state extends
almost over the whole Brillouin zone. To see this, we 
numerically study topological 
properties of bismuth ultrathin films and their edge states. 
Among bismuth thin films,
only two thin films are proposed to be insulating in the bulk:
the (111) single (1) -bilayer film~\cite{Koroteev08} and the \{012\} 2-monolayer film \cite{Yaginuma08}.
By using tight-binding Hamiltonians obtained by first-principles calculations,
we found that (111) 1-bilayer film is in the QSH phase and \{012\} 2-monolayer film is not. 
We also found that the edge states in Bi (111) 1-bilayer film are
well localized near the edges, compared with the HgTe quantum well. 
From these studies we conclude that the penetration depth $\ell$ 
corresponds to the inverse of the $k$-space width of 
edge-state dispersion.

\begin{figure}[htb]
\includegraphics[width=8cm]{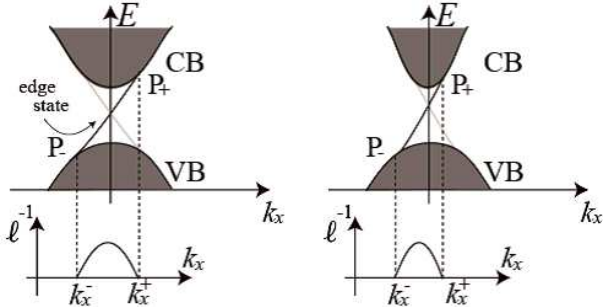}
\caption{\label{fig:penetration3}
Penetration depth $\ell$ for the effective model with ribbon geometry. 
CB (VB) represents the bulk conduction (valence) band. The plot 
on the right corresponds to a more asymmetric situation, leading to
a larger $\ell$ at the crossing point of the edge states.}
\end{figure}

{\it Penetration depth of the edge states}--
We use the Hamiltonian for the HgTe quantum well.
\begin{equation}
\mathcal{H}\left( k_{x},k_{y}\right) =\left( 
\begin{array}{cc}
H\left( {\bf k}\right) & 0 \\ 
0 & H^{\ast }\left( -{\bf k}\right)%
\end{array}%
\right) ,  \label{H}
\end{equation}%
where $H\left( {\bf k}\right) =\epsilon _{{\bf k}}
\mathbf{I}_{2}+d^{a}\left( {\bf k}\right)
\sigma ^{a}$. Here, $\mathbf{I}_{2}$ is a $2\times 2$ unit matrix, $\sigma
_{a}$ the Pauli matrices, $\epsilon _{{\bf k}}=C-D\left(
k_{x}^{2}+k_{y}^{2}\right) $, $d^{1}=Ak_{x}$, $d^{2}=Ak_{y}$, and $d^{3}=%
\mathcal{M}\left( k\right) =M-B\left( k_{x}^{2}+k_{y}^{2}\right) $. 
The constant $C$ is set to zero since it is an overall energy offset. 
The eigenenergies are then given by $-Dk^2\pm|{\bf d}({\bf k})|$.
Thus $D$ represents the asymmetry between the valence and the conduction 
band  dispersions . 
The bulk gap at ${\bf k}=0$ is given by $2M$.
In order to consider the edge state on a single edge, 
we deal with a system on a half-plane
of $y\leq 0$. This considerably simplifies the results, compared with 
the ribbon of finite width \cite{Zhou08}.
As edge states only the solutions with 
$e^{\lambda y}$ (${\rm Re}\lambda>0$) are allowed. 
The secular equation
\begin{eqnarray}
&&(M-E+B_+
(\lambda^2-k_x^2))(-M-E-B_-(\lambda^2-k_x^2))\nonumber\\
&&\ \ \ \ =A^2(k_x^2-\lambda^2)
\end{eqnarray}
where $B_{\pm }=B\pm D$,
gives two allowed values for $\lambda$:
\begin{equation}
\lambda=\lambda _{1,2}=\sqrt{k_{x}^{2}+F\pm \sqrt{F^{2}-
(M^{2}-E^{2})/(B_{+}B_{-})}},
\label{LK}
\end{equation}%
where $F=\frac{A^{2}-2\left( MB+ED\right) }{2B_{+}B_{-}}.$
If we impose a boundary condition $\psi(y=0)=0$ as in \cite{Zhou08}, 
we get 
\begin{equation}
\lambda_1\lambda_2=\frac{BM+DE}{B_+B_-}-k_x^2, \ \ 
\lambda_1+\lambda_2=\frac{DM+BE}{k_xB_+B_-}\label{sum}.
\end{equation}
From Eqs.~(\ref{LK}) and (\ref{sum}), 
we obtain an exact form for the dispersion of edge states
\begin{equation}
E=-\frac{DM}{B}\pm\frac{A}{B}\sqrt{B_+B_-}k_x.
\label{dispersion}
\end{equation}
The signs correspond to the two branches of edge states with 
opposite spins. Because 
they are related with each other by Kramers theorem, we henceforth 
consider only the plus sign in (\ref{dispersion}).
Putting (\ref{dispersion}) into (\ref{sum}), we get
\begin{equation}
\lambda_1\lambda_2=-k_x^2+\frac{2DN}{B}k_x+\frac{M}{B},\ 
\lambda_1+\lambda_2=2N,
\label{sum2}
\end{equation}
where $N=A/(2\sqrt{B_+B_-})$.
These determine $\lambda_{1,2}$. If we put $\lambda_1>\lambda_2$, 
$\lambda_2^{-1}$ gives the physical penetration depth $\ell$ 
as discussed in 
\cite{Zhou08}. 
At the points with $\lambda_2=0$, the edge states have infinite 
penetration depth and become bulk states. From (\ref{sum2}) 
this occurs when
$
k_x=k_x^\pm\equiv
\frac{DN}{B}\left(1\pm \sqrt{1+
\frac{BM}{D^2N^2}}\right)$.
It can be checked that the states at $k_x=k_x^{\pm}$ are located 
at the band edge of the 
projection of the bulk band, and at these points the edge dispersion
(\ref{dispersion}) is tangential to the bulk band projection.
We can rewrite as
$\lambda_1\lambda_2=-(k_x-k_x^+)(k_x-k_x^-)$. Therefore $\lambda_2(=\ell^{-1})$ is expressed as
$\ell^{-1}=N-\sqrt{N^2+(k_x-k_x^+)(k_x-k_x^-)}$. 
Hence the behavior of $\ell^{-1}$ is as shown in Fig.~\ref{fig:penetration3}.
It vanishes at the points $P_{\pm}$ ($k_x=k_x^{\pm}$) where the edge states are absorbed into 
the bulk band, and $\ell$ is minimum when $k_x=(k_x^{+}+k_x^{-})/2$.
The minimum value $\ell_{\rm min}$ is given by
$
\ell_{\rm min}^{-1}=
N-\sqrt{N^2-
(k_x^+-k_x^-)^2/4}$.
As a function of $N$, the minimum value of 
$\ell_{\rm min}$ is $2/(k_x^+-k_x^-)$ at $N=(k_x^+-k_x^-)/2$. 
This means that the minimum $\ell_{\rm min}$ of the system 
 is roughly given by the inverse of the $k$-space extension of the edge
state dispersion.
From Fig.~\ref{fig:penetration3} it can be seen that 
the penetration depth $\ell$ becomes short
when the considered  edge state is far from the points $P_{\pm}$. 
The inverse of 
the penetration depth $\ell^{-1}$ 
corresponds to an imaginary part of the wavenumber perpendicular
to the edge direction, and therefore it behaves similarly 
to the (real) wavenumber. 
Hence $\ell^{-1}$ is approximately given by the
$k$-space distance of wavenumbers from the points $P_{\pm}$.

In the HgTe quantum wells, the 2D quantum spin Hall states 
are confirmed by transport measurements \cite{Konig07,Roth09}. 
The penetration depth of the edge states in these systems has been 
calculated to be relatively long $\ell \sim 50{\rm nm}$ \cite{Zhou08}.
In our theory, by plugging the parameters into our results, we get $\ell=56$nm at $k_x=0$, in agreement with \cite{Zhou08}. 
The coefficient $D$ gives rise to an asymmetry 
between the conduction and the valence bands, and 
the edge state is also asymmetric: $k_x^+\neq -k_x^-$, 
$k_x^+=0.62$nm$^{-1}$, 
$k_x^-=-0.024$nm$^{-1}$.
Thus the penetration depth $\ell$ is shortest not at $k_x=0$ but 
at $k_x=(k_x^++k_x^-)/2=0.30$nm$^{-1}$ with $\ell_{\rm min}\sim 6.2$nm.
In our interpretation, the relatively 
long $\ell$ of the edge states in HgTe quantum
well comes from the fact that the edge states are localized within a very 
narrow region in $k$ space, 
giving a long $\ell$.
This penetration depth determines the 
minimal width of the system size required for observation of 
edge states.

{\it Bi(111) Ultrathin Film}--
By extending our theory to generic types of edge states, we can 
expect that the inverse of the penetration depth $\ell^{-1}$ well 
scales with the $k$-space distance from the absorption point $P_{\pm}$ into 
the bulk band. Therefore, if the edge states extend over the Brillouin 
zone, the penetration depth of the edge states is as short as a few lattice
constants. We will theoretically show that Bi(111) ultrathin film is a QSH system having edge states with such a short penetration depth.

For the calculation, we use a tight-binding model
constructed from maximally localized Wannier orbitals \cite{Marzari97} obtained 
from first-principle calculations \cite{Freimuth08}.
The Fermi energy lies in the 6p-like states,
comprising three conduction bands and three valence bands. 
Therefore, in constructing the 
Wannier orbitals we only retain these six bands. 
From these Wannier orbitals including
the lattice relaxation effects of the ultrathin films,
we construct tight-binding models keeping 
up to third-neighbor hopping amplitudes.

\begin{figure}[htb]
\includegraphics[width=\hsize]{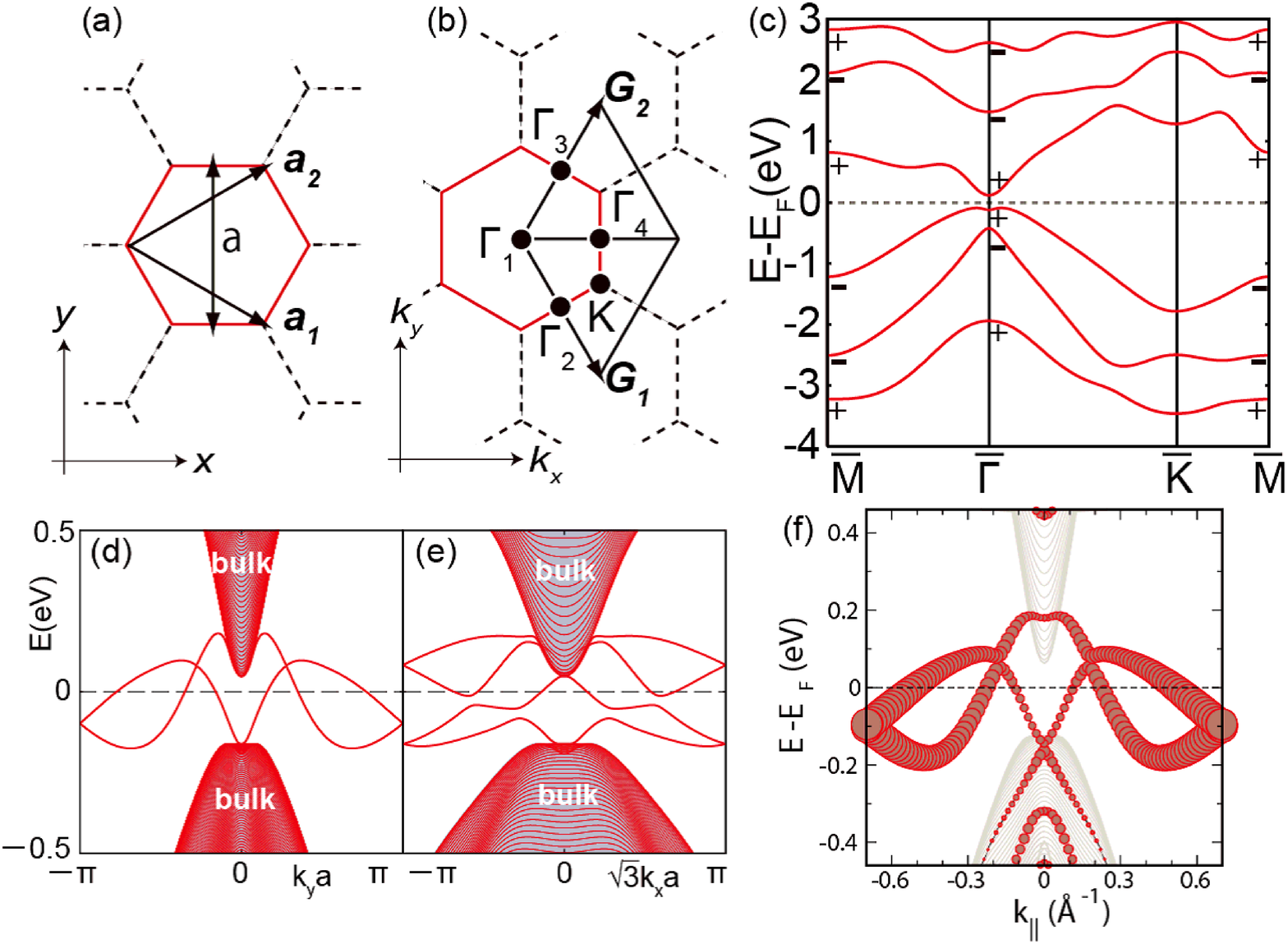}
\caption{\label{fig:bi111}
(a) Unit cell and lattice vectors, and (b) TRIMs 
of Bi(111) ultrathin film.
The TRIM $\Gamma_1$ is the $\Gamma$ point, and the three TRIMs
$\Gamma_2,\Gamma_3$ and $\Gamma_4$ are the $M$ points.
(c) Bulk energy bands and the parity at the TRIMs for a Bi(111) 1-bilayer. 
(d) and (e): Energy bands of the Bi(111) zigzag and armchair edge ribbons, respectively, 
with a width of 20 unit cells, calculated from tight-binding model. 
(f) Energy bands of a eight-unit-cell wide Bi(111) zigzag edge ribbon from
first-principles calculations. The size of the symbols corresponds to
the weight of the states in the edge atoms.}
\end{figure}

Figures~\ref{fig:bi111} (a -- c) show the unit cell and lattice vectors, 
reciprocal vectors and TRIMs, and the energy band of Bi(111) 1-bilayer respectively.
Since this system is inversion symmetric, all states in 
Fig.~\ref{fig:bi111} (c)
are doubly degenerate. 
This system is proposed to be a nonmagnetic insulator with a bulk gap 
of $\sim 0.2$eV \cite{Koroteev08}. 
We will calculate the $Z_2$ topological number $\nu$, 
and if it is nontrivial it is a QSH insulator.
The wavenumbers satisfying $\mathbf{k}\equiv -\mathbf{k}$ $({\rm mod}~\mathbf{G})$ are
called time-reversal-invariant momenta (TRIM) 
$\mathbf{k}=\mathbf{\Gamma}_{i}$ $(i=1,2,3,4)$. 
For inversion-symmetric systems, the $Z_2$ topological number $\nu$ is defined 
by 
$(-1)^{\nu}\equiv\prod_{i=1}^{4} \prod_{m=1}^{n}
\xi_{2m}({\bf \Gamma}_i)$ ($=\pm 1$),
where 
$\xi_{2m}({\bf \Gamma}_i)$ ($=\pm 1$) is the parity eigenvalue of the 
Kramers pairs at ${\bf \Gamma}_i$ and $n$ is the number of the Kramers 
pairs of eigenstates below the Fermi energy \cite{Fu07b}. 
The parity eigenvalues 
at the TRIMs $\Gamma_i(i=1,2,3,4)$ 
are given in Fig.~\ref{fig:bi111} (c) and  yield
 the topological number $\nu=1$.
We note that both the first-principle calculation 
(without a tight-binding model)
and the calculation of the Liu-Allen tight-binding model \cite{Liu95} 
 give $\nu=1$. 
In Ref.~\cite{Murakami06a} a
(111) 1-bilayer bismuth is proposed to be in the QSH phase,
from a 
simple truncation of the 3D tight-binding model \cite{Liu95}.
We thus confirmed that 
the conclusion in Ref.~\cite{Murakami06a} remains unaltered in 
first-principle calculations.

If we neglect the out-of-plane coordinate, 
the (111) 1-bilayer film has a honeycomb structure.
Therefore, as in graphene
we refer to the two types of simple edge shapes as zigzag and armchair edges. 
Figure~\ref{fig:bi111} (d)(e) shows the energy bands of zigzag and armchair edge ribbons
of the Bi(111) 1-bilayer. 
Due to inversion symmetry, all the states are doubly degenerate, 
and they have opposite spins, localized on the opposite edges.
In both figures, the number of Kramers pairs of edge states on the Fermi energy per one edge is odd,
confirming that 
Bi(111) 1-bilayer is a QSH system. 
We checked that for the zigzag-edge ribbon 
our result from the tight-binding model 
(Fig.~\ref{fig:bi111} (d)) 
and that from a first-principle calculation 
(Fig.~\ref{fig:bi111} (f)) are in good agreement.

These edge states are quite different from those in a HgTe quantum well, where the
edge states exist only near the $k=0$ point 
\cite{Bernevig07a}. 
Within our calculation, the edge states extend almost all over the whole Brillouin
zone. At the Fermi energy there are three Kramers pairs of edge states.
Thus, the conductance in a ribbon geometry becomes $G=6e^2/h$ for a
clean system. 
When nonmagnetic disorder is increased, some of these edge states 
become gapped due to elastic scattering, while at least one pair of edge states remain gapless, 
giving the conductance of $G=2e^2/h$.
These edge states form perfectly conducting channels, 
similar to those in the graphene nanoribbons \cite{Wakabayashi07}.
In graphene, perfectly conducting channels are formed 
only in the absence of short-ranged disorder; 
in the Bi (111) 1-bilayer nanoribbon the perfectly conducting 
channel exists irrespective of the nature of nonmagnetic 
disorder, and it gives
a universal behavior realizable in experiments.

{\it Bi\{012\} ultrathin film}--
For inversion asymmetric systems such as Bi \{012\}
2-monolayer film, the 
calculation of $\nu$
is complicated because the phases of the Bloch wavefunctions
in the entire Brillouin zone are involved
\cite{Kane05a,Fu06a}. 
The phase
of the wavefunction is a gauge degree
of freedom and can be chosen arbitrary for each ${\bf k}$.
Hence, a simple discretization of a formula for 
continuous ${\bf k}$ suffers from numerical instability due to this gauge choice. 

Hence, we  adopt a gauge-invariant discretization method proposed in 
Ref.~\cite{Fukui07}.
It is a merit of the method that 
we do not need to determine the phase of the wavefunction smoothly in 
${\bf k}$ space.
The mesh size $\delta k_1 \delta k_2$ should be fine enough to
satisfy 
$\left| F({\bf k}) \right| \delta k_1 \delta k_2 < \pi$
at any mesh, where 
$F({\bf k})$ is the Berry curvature,
and $\delta k_1, \delta k_2$ are the width and height of a mesh, 
respectively \cite{Fukui07}.
This quantity is largest when ${\bf k}$ is at the direct gap 
${\bf k}={\bf k}_{\rm g}$,  
and the critical size is approximated by the ${\bf k}$-space nominal size of the 
band extremum at ${\bf k}={\bf k}_{\rm g}$.
From the band structure of Bi\{012\} 2-monolayer, 
the critical mesh number $n_B^c$ is estimated to be $\sim 100$.
For various mesh numbers exceeding $n_B^c$  we get the consistent 
result that the $Z_2$ topological number is $\nu=0$. Therefore, Bi\{012\} 2-monolayer is an ordinary insulator.

\begin{figure}[htb]
\includegraphics[width=8cm]{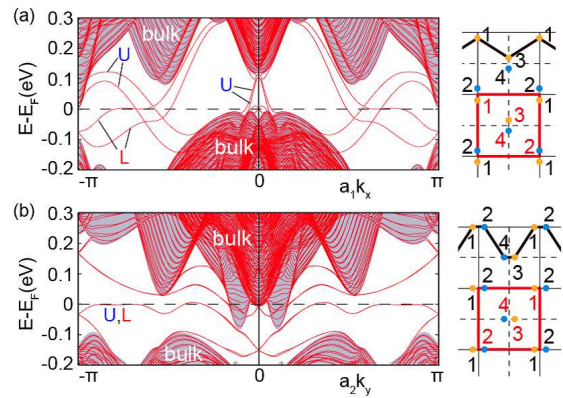}
\caption{\label{fig:bi012_ribbon_energyband}
Energy bands of Bi\{012\}: (a) zigzag and (b) armchair edge ribbons
with a width of 20 unit cells.
`U' (`L') means that the state is localized on the upper (lower) edge.
The crystal structures near the upper edge 
are shown in the right panels 
with 1,2,3 and 4 representing the lattice sites.
We note that these 4 sites 
do not lie on the same plane. The shaded regions are the bulk energy bands.
}
\end{figure}

The edge states for 
ribbons with 
two types of edges of Bi\{012\} 2-monolayer ribbons and their energy bands
are shown in Fig.~\ref{fig:bi012_ribbon_energyband}.
The two edge shapes can be called zigzag and 
armchair edges, although the lattice structure 
is quite different from graphene.  
The number of Kramers pairs 
of edge states on the Fermi energy is even at each edge
and this is in agreement with our result that the $Z_2$ topological number is 
$\nu=0$. In the armchair-edge ribbon, 
the edge states are almost degenerate because of the equivalence of two edges
of the ribbon via mirror-symmetry.
These two states have an energy difference,
due to hybridization of the edge states at the opposite edges.
Nevertheless, for a ribbon wider than $\ell$, the energy difference is 
exponentially small.
On the contrary, in the zigzag-edge ribbon, 
the edge states are not degenerate because of
inequivalence between both edges of the ribbon.

\begin{figure}[htb]
\includegraphics[width=8cm]{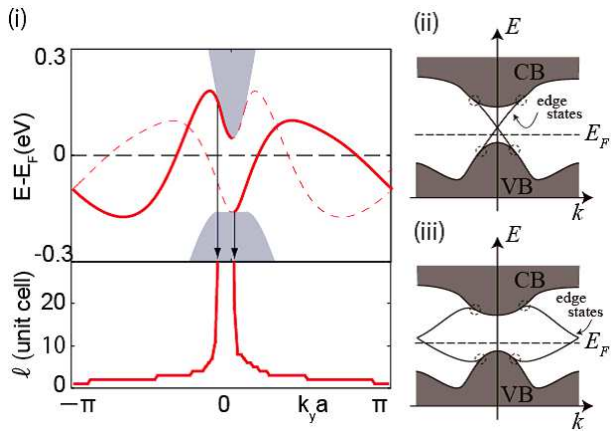}
\caption{\label{fig:locallength}
(i) Penetration depth of the edge states on the zigzag 
edge of the Bi(111). (ii)(iii) Examples of the edge states in the 
2D QSH systems. 
}
\end{figure}

{\it Penetration depth of the edge states}--
We now calculate the penetration depth $\ell$ of the edge states 
of the Bi(111) 1-bilayer film. 
For the edge states on the zigzag edge of the (111) 1-bilayer film, 
the result is shown in Fig.~\ref{fig:locallength} (i).
The penetration depths $\ell$ of the edge states (including those on the
Fermi energy) are typically 
several lattice constants.
Hence, for transport experiments
the width of the sample has  to be larger than a few lattice constants.

These results on Bi(111) film agree with our theory on $\ell$. 
According to our theory, 
the penetration depth $\ell$ becomes short
when the edge states are distant from the points $P_{\pm}$ where the edge states merge into the bulk (circles in Fig.~\ref{fig:locallength}(ii)(iii)).
Hence, $\ell$ is longer for the states at $E_{\rm F}$ in
Fig.~\ref{fig:locallength} (ii), and shorter in Fig.~\ref{fig:locallength} (iii).
 This information is relevant for transport which is governed by the states at the Fermi level.  
In Bi(111), the edge state 
travels almost over the whole Brillouin zone (BZ). Therefore we 
estimate $\ell\sim \mbox{(size of the BZ)}^{-1}\sim \mbox{(lattice 
spacing)}$, in agreement with the results in Fig.~\ref{fig:locallength}(i)

Bi(111) 1-bilayer film cannot be described by an effective model
near ${\bf k}=0$ like (\ref{H}). The effective model (\ref{H}) is 
derived when the QSH system is described as a band inversion between 
two doubly-degenerate bands, such as HgTe quantum well, or Bi$_2$Se$_3$. 
In bismuth ultrathin 
films, the involved bands are $p_x$, $p_y$, $p_z$ orbitals, and the valence and 
conduction bands have different mixing coefficients for these orbitals. 
Therefore, it is not a mere band inversion, which is the 
reason why the case Fig.~\ref{fig:locallength}(iii) is realized in 
bismuth films.
Bi$_2$Te$_3$ and Bi$_2$Se$_3$ ultrathin films also have edge states similar
to Fig.~\ref{fig:locallength}(iii) \cite{Liu09}. 
In these cases, however, some 
edge states are close to the bulk bands, leading to 
very long penetration depths of about a hundred times the lattice constant .
 We note that our theory assumes isotropy between the direction along the edge/surface
and that perpendicular to it. For layered materials such as ${\rm Bi_2Se_3}$ and ${\rm Bi_2Te_3}$, 
the penetration depth perpendicular to the layer cannnot be predicted
from the surface-state dispersion in the layer because of the anisotropy. 

These short penetration depths of edge states in the Bi (111) 1-bilayer film are
ideal for observation by STM/STS and 
control of the edge states.
Furthermore, it is also favorable 
for edge thermoelectric transport \cite{Takahashi09}.
To utilize the perfectly conducting channels of edge states for
thermoelectric transport, short penetration depth is an important
factor, because longer penetration depth mixes the 
states at different edges for narrow ribbons, and 
destroys the coherent electron transport at the edges.

{\it Conclusion}--
We analyze a generic behavior of the penetration depth of the edge states
 in two-dimensional quantum spin Hall systems. 
We found that momentum-space
distance between the edge states and the absorption point 
of the edge dispersion into the bulk band roughly gives the inverse of the 
penetration depth. 
As an example, we calculate the penetration depth 
of the edge states of Bi(111) 1-bilayer film, which we propose to 
be a QSH insulator. 
The penetration depth of the edge states in Bi(111) 1-bilayer film 
is in good agreement with our theory.
 
We are grateful to S. Bl\"ugel, T. Hirahara, 
T. Nagao, and S. Yaginuma for helpful discussions. 
This research is supported in part 
by MEXT KAKENHI.  

\end{document}